\documentclass[twocolumn, final]{svjour3}
\usepackage{graphicx}
\usepackage{latexsym}
\usepackage{color}
\usepackage{makeidx}

\begin{document}

\authorrunning{T. Gubiec and R. Kutner}
\titlerunning{Infinite-step memory within the CTRW ...}
\title{Continuous-Time Random Walk with multi-step memory: An application to market dynamics}
\author{Tomasz Gubiec \and Ryszard Kutner} 
\institute{Faculty of Physics, University of Warsaw, Pasteur Str. 5, PL-02093 Warsaw, Poland}
\date{Received: date / Revised version: date}





\maketitle

\begin{abstract}
A novel version of the Continuous-Time Random Walk (CTRW) model with memory is developed.
This memory means the dependence between arbitrary number of successive jumps of the process, while waiting times between jumps are considered as i.i.d. random variables. 
The dependence was found by analysis of empirical histograms for the stochastic process of a single share price on a market within the high frequency time scale, and justified theoretically by considering bid-ask bounce mechanism containing some delay characteristic for any double-auction market.
Our model turns out to be exactly analytically solvable, which enables a direct comparison of its predictions with their empirical counterparts, for instance, with empirical velocity autocorrelation function. Thus this paper significantly extends the capabilities of the CTRW formalism.
\end{abstract}

\newcommand {\mathsym}[1] {{}}
\newcommand {\unicode} {{}}
\newcommand{\ho}{\right|_{h=0}}
\newcommand{\hh}{\left.}
\newcommand{\bee}{\begin{eqnarray}}
\newcommand{\eee}{\end{eqnarray}}
\newcommand{\be}{\begin{equation}}
\newcommand{\ee}{\end{equation}}
\newcommand{\bifi}{\begin{figure}}
\newcommand{\efi}{\end{figure}}
\newcommand{\Q}[2]{Q_{#1} \left( #2 \right)}
\newcommand{\tQ}[2]{\tilde{Q}_{#1} \left( #2 \right)}
\newcommand{\ttQ}[2]{\tilde{\tilde{Q}}_{#1} \left( #2 \right)}
\newcommand{\psit}{\psi(t)}
\newcommand{\psitf}{\psi_1(t)}
\newcommand{\tpsi}{\tilde{\psi}}
\newcommand{\psis}{\tilde{\psi}(s)}
\newcommand{\psisj}{\tilde{\psi}^j(s)}
\newcommand{\psisf}{\tilde{\psi}_1 (s)}
\newcommand{\epsis}{\left[ \epsilon \tilde{\psi}(s)\right]}
\newcommand{\epsisn}{ \epsilon \tilde{\psi}(s)}
\newcommand{\tilh}{\hat{h}}
\newcommand{\hk}{\hat{h}(k)}
\newcommand{\tsr}{\left\langle \tau \right\rangle}
\newcommand{\eiw}[1]{e ^{i k #1}}
\newcommand{\ip}[1]{\int\limits^{\infty}_{0} d #1}
\newcommand{\im}[1]{\int\limits^{0}_{-\infty} d #1}
\newcommand{\iall}[1]{\int\limits^{\infty}_{-\infty} d #1}
\newcommand{\edo}[1]{e ^{#1} \,}
\newcommand{\AAA}{\frac{1-2 \epsilon}{1-\epsilon}}
\newcommand{\BB}{\frac{\zeta}{1-\epsilon}}
\newcommand{\CC}{\frac{\epsilon-\zeta}{1-\epsilon}}
\newcommand{\DD}{\frac{\epsilon-\zeta}{\epsilon}}
\newcommand{\EE}{\frac{\zeta}{\epsilon}}
\newcommand{\refe}[1]{(\ref{#1})}
\newcommand{\tQS}[2]{\tilde{Q}^S_{#1} \left( #2 \right)}
\newcommand{\tQR}[2]{\tilde{Q}^R_{#1} \left( #2 \right)}

\section{Introduction}

The dynamics of many complex systems, not only in natural but also in socio-economical sciences,
is usually represented by stochastic time series. These series are often composed of elementary random spatio-temporal events, which may show some dependences and correlations as well as apparent 
universal structures~\cite{barabasi2005,vazquez2007,yamasaki2005,nakamura2007,vazquez2006,perello2006}. 
By this elementary event we understand a "spatial" jump, $r$, of a stochastic process preceded by waiting (interevent or pausing) 
time, $\tau$, both being stochastic variables. 

Such a two-phase stochastic process, named Continuous-Time Random Walk (CTRW), was introduced in the physical context of dispersive transport and diffusion by Montroll and Weiss \cite{montroll1965} and applied successfully to description of a photocurrent relaxation in amorphous films \cite{SM,pfister1978,shlesinger1984,weiss1994,bouchaud1990} (and ref. therein) and in OLED ones \cite{gill1972,campbell1997}.

The CTRW formalism was applied for example, for diffusion in probabilistic fractal structures such as percolation clusters \cite{ben2000} and for fractional diffusion \cite{hilfer1999}.
The CTRW with broad waiting time distribution was applied, e.g., for diffusion in chaotic systems \cite{geisel1995}.
The CTRW formalism, containing broad spatial jump distribution explained superdiffusion (L\'{e}vy flights or walks) \cite{klafter1995} observed in domains of rotating flows or weakly turbulent flow \cite{weeks1995,weeks1998}.
The CTRW found innumerable applications in many other fields: hydrogen diffusion in nanostructure compounds \cite{hempelmann1999},
nearly constant dielectric loss in disordered ionic conductors \cite{dieterich2009}
subsurface tracer diffusion \cite{scher2002},
electron transfer \cite{nelson1999},
aging of glasses \cite{EB,MB},
transport in porous media \cite{margolin2000},
diffusion of epicenters of earthquakes aftershocks \cite{helmstetter2002},
cardiological rhythms \cite{iyengar1996},
search models \cite{lomholt2008},
human travel \cite{hufnagel2006}
and even financial markets \cite{f6,kutner2002,scalas2006,perello2008,kasprzak2010}.
Today, the CTRW provides an unified description for both enhanced and dispersive diffusion \cite{kutner1999a,kutner1999b,metzler2000,metzler2004} - the list of its applications is still growing (cf. \cite{arXivEPJB}).

Nearly three and a half decades ago the versions of the CTRW formalism containing the backward or forward correlations between jump directions were developed \cite{haus1987} (and refs. therein). 
Soon, the first application of the former version of the formalism,  as in the case of concentrated lattice gas, was performed for the study of the tracer diffusion coefficient \cite{kehr1981}.
The study was directly inspired by hydrogen diffusion in transition metals \cite{springer1972,kutner1977} and ionic conductivity in super-ionic conductors \cite{salamon1979}.
As a result, for lattices of low coordination numbers or networks with low average nodes' degrees, the description of the tracer diffusion in concentrated lattice gas requires an extension of the CTRW formalism to take into account the dependences over several subsequent jumps \cite{kutner1985}. 
This can occur because the vacancy left behind the tracer particle after its jump favorizes the return of the tracer to the origin, even after several jumps.
The CTRW formalism with memory appeared also in other contexts \cite{montero2007,montero2011}, but up to now, still limited only to the dependence over two subsequent jumps as its extension to the case of memory (or dependence) over three or more subsequent jumps was too complicated for the theoretical derivation.

This work extends the field of applications of the CTRW formalism by including memory ranging over two jumps behind the current jump. 
In other words, in this work the dependence between \underline{three} subsequent jumps is considered resulting in an exact analytical solution.
Such an approach is useful not only for study of one dimensional random walk but also can be useful in higher dimensions for different kinds of lattices and networks.

Furthermore, we applied our CTRW formalism to the subtle description of the high-frequency price dynamics driven by the microscopic mechanism of bid-ask bounce phenomena.
 One reason in favor of CTRW formalisms is that they provide a generic formula for the first and second order time-dependent statistics in terms of two auxiliary spatial, $h(r)$, and temporal, $\psi(\tau)$, distributions that can be obtained directly from empirical histograms.

The paper is organized as follows: in Sec. \ref{sec:mot} we present the motivation of our work. 
In Sec. \ref{sec:def} we define the proper stochastic process which is solved in the Sec. \ref{sec:sol}. 
In Sec. \ref{sec:com} the novel model is compared with our previous model \cite{gubiec2} and in Sec. \ref{sec:emp} the comparison with empirical data was made. 
Section \ref{sec:sum} contains our concluding remarks.

\section{Direct motivation} \label{sec:mot}

There are few (considered below) direct reasons supplied, for instance, by the financial markets, which pushed us to include the two-step memory into the Continuous-Time Random Walk formalism in a generic way.
{

If we record for simplicity only successive share price jumps and not time intervals (waiting-times) between them, we obtain the so-called ``event-time'' series. 
The event-time dependent autocorrelation functions of price changes obtained on this basis were already widely considered \cite{dacorogna2001,tsay}. 
The shape of these autocorrelation functions, that is, their dependence on event-time is universal in the sense that the shape is independent of the market and stock analyzed and for each considered event-time series we get the distinctly negative value of lag-1 autocorrelation function, while almost vanishing values for lag-2, lag-3, $\ldots$. For this reason, the shape of this autocorrelation function can be considered as a stylized fact.

The significant correlation between two successive price jumps stimulated both Montero and Masoliver \cite{montero2007} as well as authors of the present work \cite{gubiec2} to describe the stochastic process of the single stock price as a CTRW with one step backward memory, in which current value of the increment depends only on the previous one. Such a dependence is caused in finance by the bid-ask bounce phenomenon \cite{roll,dacorogna2001}. 
Previously we assumed for simplicity \cite{gubiec2} that dependence between current price jump and the second one before the current price jump can be neglected as corresponding correlation vanishes.
However, in the present work the mentioned above dependence is taken into account as we observed, herein, that even vanishing of the correlation does not imply the lack of dependence -- this is a key obervation which initialized the present work.

We remind that by basing on the empirical histogram of the two consecutive price jumps (compare diagrams in Fig. 1 in ref. \cite{gubiec2} and the analogous one in Fig. \ref{two:f:dx1dx3}a in the present work), we proposed a formula which describes dependence (herein of the backward form) between two consecutive (lag-1) jumps, $r_n, r_{n-1}$, by the joint two-variable pdf
\bee
h (r_n, r_{n-1})&=&(1 - \epsilon )  h(r_{n}) h(r_{n-1}) \nonumber \\
&+&\epsilon \ \delta(r_{n}+r_{n-1}) h(r_{n-1}), \label{eq:hold}
\eee
or equivalently by the conditional pdf
\bee
h (r_n \mid r_{n-1})= (1 - \epsilon )  h(r_{n}) + \epsilon \ \delta(r_{n}+r_{n-1}), \label{eq:hmidold}
\eee
where $h(x)$ is an even function as no drift is present herein and $0\leq \epsilon \leq 1$ is a constant weight, which can be estimated either from the histogram or from the lag-1 autocorrelation function of consecutive jumps of the process. 
Apparently, only the second term in Eqs. \refe{eq:hold} and \refe{eq:hmidold} describes dependence (herein of the backward type) between $r_n$ and $r_{n-1}$ variables.
Furthermore, above formulas imply a dependence between $r_n$ and $r_{n-2}$ jumps, expressed in the two-variable pdf
\bee
h_2 (r_n, r_{n-2})&=& \iall{r_{n-1}}  h (r_n \mid r_{n-1}) h (r_{n-1}, r_{n-2})=\nonumber \\
&=& (1 - \epsilon^2 )  h(r_{n}) h(r_{n-2}) \nonumber \\
&+& \epsilon^2 \ \delta(r_{n}-r_{n-2}) h(r_{n-2}),
\label{eq:h2wrong}
\eee
which gives a significant, positive correlation between $r_n$ and $r_{n-2}$ equals $\epsilon^2$.

The generalization of Eq. \refe{eq:h2wrong} for the dependence between any two jumps is straightforward
\bee
h_k (r_n, r_{n-k})&=& \iall{r_{n-1}} \ldots \iall{r_{n-k+1}}  h (r_n \mid r_{n-1}) \nonumber \\
&\times& h (r_{n-1} \mid r_{n-2}) \ldots h (r_{n-k+1}, r_{n-k})=\nonumber \\
&=& (1 - \epsilon^k )  h(r_{n}) h(r_{n-k}) \nonumber \\
&+& \epsilon^k \ \delta(r_{n}-(-1)^k r_{n-k}) h(r_{n-k}), \nonumber \\
& &k=2,3, \ldots. 
\label{eq:hkwrong}
\eee
where $k$ is the number of steps in the event time.
Hence, the autocorrelation function of jumps in the event time is simply
\bee
c(k)=\frac{1}{\mu_2} \iall{r_{n}}\iall{r_{n-k}} r_{n} r_{n-k} h_k (r_n, r_{n-k})=(-\epsilon)^k, \nonumber \\
\label{eq:ckwrong}
\eee
where the second moment $\mu_2=\iall{x} \ x^2 \ h(x)$.
However, relation \refe{eq:ckwrong} is not observed in empirical data as empirical autocorrelation function decreases to zero much quicker. 

In principle, the empirical autocorrelation function between jumps cannot be reproduced if one assumes that (i) only two successive jumps are dependent and (ii) this dependence is described by the symmetric distribution function $h(r_n, r_{n-1})=h(r_{n-1},r_n)$, although the latter is justified by the empirical representation of $h(r_n, r_{n-1})$ shown in Figure \ref{two:f:dx1dx3}a.
As a consequence of assumption (i) the correlation between $r_n$ and $r_{n-2}$ is always greater than zero (see Appendix \ref{section:Proof} for detailed derivation), which essentially disagrees with empirical data shown in Figure \ref{two:f:dx1dx3}b.
Indeed, this disagreement is one of the main inspirations to consider the CTRW model with longer memory, where each current jump of the process depends on the two previous jumps, supplying an exact analytical solution.

\section{Definition of the model}\label{sec:def}

Let us begin with the analysis of the empirical histogram presenting dependence between the current price jump and the second one before the current jump. 
This histogram, which is a statistical realization of the function $h_2 (r_n, r_{n-2})$, is shown in Figure \ref{two:f:dx1dx3}b.
\begin{figure}[ht]
\centering
		\includegraphics[width=0.49\textwidth]{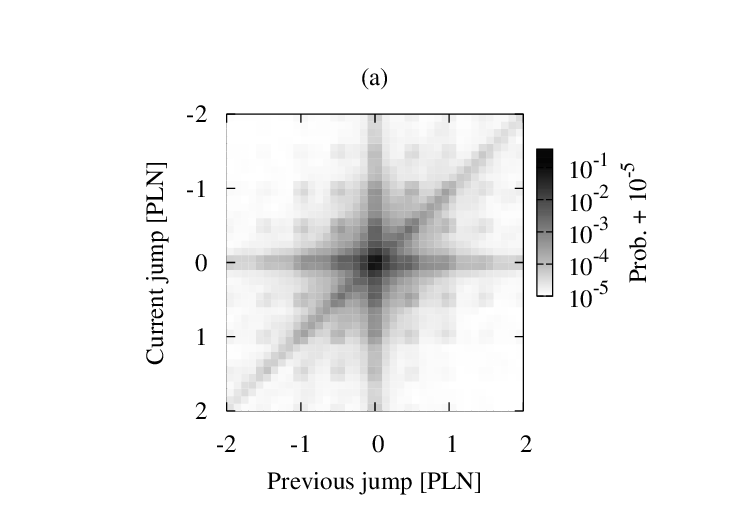}
 		\includegraphics[width=0.49\textwidth]{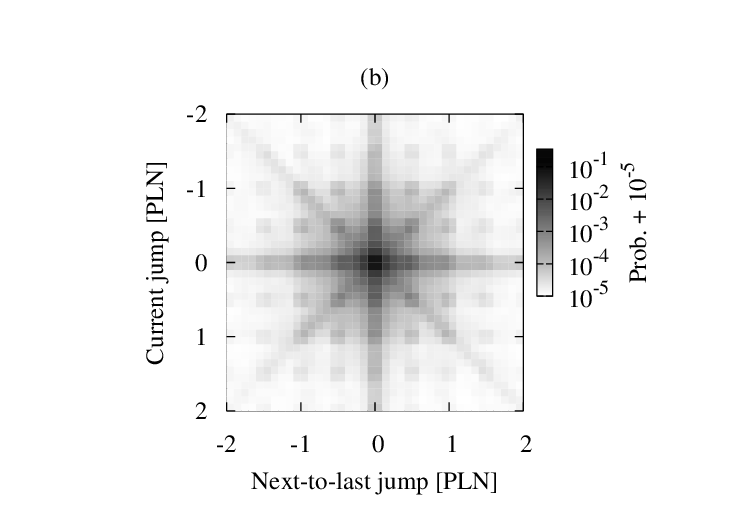}	
	\caption{Empirical normalized histograms of different kinds of two price jumps dependences: (a) for the current price jump and the preceding price jump, (b) for the current price jump and the second one before the current price jump or next-to-last jump. 
The larger logarithm of the joint	probability is visualized by more intense grayness. 
To avoid singularity of the logarithmic scale, all probabilities are increased by small insignificant number $10^{-5}$.}\label{two:f:dx1dx3}
\end{figure}
Observed antisymmetric dependence between $r_n$ and  $r_{n-2}$ can be considered as a generic empirical example of two random variables which are dependent but uncorrelated. 
Besides a sharp central cross, it contains both a ``diagonal'' and an ``anti-diagonal''. 
These diagonals and anti-diagonals correspond to the case, where the current price jump and the second one before the current price jump have the same length but might have the same or the opposite signs. 

Apparently, Eq. \refe{eq:h2wrong} is able to reproduce only the diagonal of the histogram. 
To reproduce both diagonal and anti-diagonal we ave to extend Eq. \refe{eq:h2wrong} into the form
\bee 
h_2 (r_n, r_{n-2}) &=& (1 - 2 \zeta )  h(r_{n}) h(r_{n-2}) \nonumber \\ 
&+& \zeta \ \delta(r_{n}-r_{n-2}) h(r_{n-2}) \nonumber\\
&+& \zeta \ \delta(r_{n}+r_{n-2}) h(r_{n-2}), \quad 0\leq\zeta\leq 1, \nonumber \\
\label{eq:h2}
\eee
essential for further considerations, where the second and third terms represent diagonal an anti-diagonal, respectively. 
These terms, together with the first term, make distribution $h_2 (r_n, r_{n-2})$ well normalized quantity.
To obtain a vanishing correlation between $r_n$ and $r_{n-2}$ we assumed weights of the diagonal and anti-diagonal equal and denoted by $\zeta$. 
Now, we can construct the three-variable pdf of three consecutive price jumps. 
For simplicity, instead of notation ($r_n, r_{n-1},r_{n-2}$) we use ($r_3,r_2,r_1$). 

The three-variables pdf, $h(r_3,r_2, r_1)$, should obey the following constrains concerning the marginal distributions:
\begin{itemize}
	\item[(a)] Firstly, distribution $h(r_3,r_2, r_1)$ integrated over any two of the three variables should reproduce, for the third variable, a single price jump distribution -- the same for all three cases. 
	The analogical constrain for two variables pdf is already satisfied by Eq. \refe{eq:hold}. 
	\item[(b)] Secondly, distribution $h(r_3,r_2, r_1)$ integrated over variable $r_1$ should reproduce two-variables pdf, $h(r_3,r_2)$, in the form of Eq. \refe{eq:hold}.
The same pdf $h(r_3,r_2, r_1)$ integrated over variable $r_3$ should also reproduce two-variables pdf, $h(r_2,r_1)$, again in the form of Eq. \refe{eq:hold}. 
	\item[(c)] However, pdf $h(r_3,r_2, r_1)$ integrated over variable $r_2$ should give pdf $h_2 (r_3, r_1)$ in the form of Eq. \refe{eq:h2}.
\end{itemize}
Hence, we propose a key formula for $h(r_3,r_2, r_1)$ in the form
\bee
h(r_3,r_2,r_1) &=& (1-2\epsilon) h(r_3)h(r_2)h(r_1) \nonumber \\
&+& \zeta \delta(r_3+r_2)\delta(r_2+r_1)h(r_1)  \nonumber\\
&+& (\epsilon -\zeta) \delta(r_3+r_2)h(r_2)h(r_1) \nonumber \\
&+& (\epsilon -\zeta) \delta(r_2+r_1)h(r_3)h(r_1) + \nonumber \\
&+& \zeta \delta(r_3+r_1)h(r_2)h(r_1), \label{eq:h321}
\eee
which satisfies all constrains mentioned above. Obviously, this form is not a unique pdf but it is the simplest one which uses only two parameters ($\epsilon$ and $\zeta$), where additionally each term has clear interpretation.

It is worth to mention that all terms shown on the right-hand side of Eq. \refe{eq:h321}, except the last one, are present, with slightly modified pre-factors, in the simple product of distributions $h(r_3 \mid r_2)$ and $h(r_2, r_1)$ defined by Eqs. \refe{eq:hold} and \refe{eq:hmidold}, respectively. 
The only new term is the last one, proportional to $\delta (r_3+r_1) h(r_2)$. 
This term describes the case, where price jump $r_1$ is followed by the second, independent price jump $r_2$ and the third price jump $r_3=-r_1$ which has the same length as jump $r_1$ but the opposite sign. The adding of such a term is due to the bid-ask bounce phenomena with delay present herein. 
We explain what is meant by the name `bid-ask bounce with delay' by using a characteristic scenario given below.

Let us consider a continuous-time double auction market organized by the order book system \cite{dacorogna2001,ohara1995,hasbrouck2006,campbell2012}.  
Let buy and sell orders be sorted according to the corresponding price limit.
The gap between buy order with the highest price limit and sell order with the lowest price limit is called the bid-ask spread} \cite{dacorogna2001,ohara1995,hasbrouck2006,campbell2012}. 
In our previous paper \cite{gubiec2} we analyzed, as a typical example, a series of orders which lead to the bouncing of the price between lower and higher border of the bid-ask spread. 
To justify the form of Eq. \refe{eq:hold}, we argued that if the price increases from the lower border of the bid-ask spread to some possibly new value of the higher border, the two cases are possible. 

In the first case, an appropriate sell order occurs, with probability $\epsilon$, and the price goes back to the vicinity of the previous price. 
This results in two consecutive price jumps of approximately the same length but opposite signs. 
In the second case, if other type of the order arrived, it leads to the elimination of the system memory present in the bid-ask spread. 
As a result, the subsequent price jump can be considered in this case as independent of the previous jump and appears with probability $1-\epsilon$. 
These two cases can be formally expressed by the two variable pdf just in the form given by Eq. \refe{eq:hold}. 
However, as we argued in the previous section, one-step memory CTRW formalism is not able to properly describe the high frequency stock market dynamics. 

Fortunately, from the second case considered above, we are able to extract the subsequent case, leading eventually to the two-step memory. 
That is, if after the first price jump the executable small volume buy order appeared, the price jump (initiated by this buy order) will also be small or even equals zero. 
In such a case, the memory of the system is still present in the bid-ask spread, because its lower border still did not move, in fact. 
Hence, the backward jump to the lower border is still possible with the price jump of approximately the same length as the second to last price jump, but with opposite sign. Analogous dependence can be present for longer series of consecutive jumps but with systematically decreasing order.
We emphasize that we do not assume that subsequent orders are independent, so our model even describes a situation where memory is present in the order flow \cite{lillo2004}

By means of pdf, the term describing such a case (of the two-step memory) can be approximated by the term proportional to $\delta (r_3+r_1) \delta(r_2) h(r_1)$. 
The first Dirac's delta is responsible for the situation where the current jump $r_3$ repeats the second one, $r_1$, before the current price jump, but with the opposite sign (i.e. $r_3=-r_1$). 
The second Dirac's delta gives the zero-length mid price jump $r_2$. 
However, to obey all three constrains (a) - (c) on marginal distributions of $h(r_3,r_2, r_1)$,  we were forced to use instead of two deltas, the last term based on the product $\delta (r_3+r_1) h(r_2) h(r_1)$. 
Let us remind that single jump distribution $h(r_2)$ is strongly concentrated at the vicinity of $r_2$ equals zero. 
Taking this term into account with appropriate weight, we thus completed our basic Eq. \refe{eq:h321}.

In our model the jumps of the process are not independent, as a current jump depends on two preceding jumps. 
Hence, the conditional pdf of the jump length $r_3$, under the condition of previous jumps $r_2$ and $r_1$, can be obtained from Eq. \refe{eq:h321} by dividing of its both sides by $h(r_2,r_1)$ given by Eq. \refe{eq:hold}. 
This leads to the useful conditional pdf
\bee
h(r_3 \mid r_2, r_1) &=& \left(1-\delta_{r_2,-r_1}\right) \nonumber \\
&\times &\left(\AAA \ h(r_3) + \BB \ \delta(r_3+r_1)\right) \nonumber \\
&+&\left(1-\delta_{r_2,-r_1}\right) \CC \ \delta(r_3+r_2) \nonumber \\
&+& \delta_{r_2,-r_1}\left(\DD  \ h(r_3) + \EE \ \delta(r_3+r_2) \right), \nonumber \\
\label{two:hh}
\eee
where the following dependences between Dirac's delta and Kronecker's delta were used
\bee
\left(1-\delta_{x,-y}\right) \delta(x+y) &=& 0, \nonumber \\
\delta_{x,-y} \delta(x+y) &=& \delta(x+y). \nonumber 
\eee
As we precisely defined dependences between consecutive jumps, we can introduce a stochastic process and derive the analytical forms of the most significant quantities such as the propagator and velocity autocorrelation function of the process.

Notably, Eqs. \refe{eq:h321} and \refe{two:hh} have generic character, which does not limit them to the local dynamics of share price only.

\section{Solution}\label{sec:sol}

The high-frequency share price time series can be considered as a single realization or trajectory of a jump stochastic process.
The trajectory of such a process is a step-way function consisting of waiting times $\tau_n$ prior to the sudden jump increment of a price $r_n$. 
Hence, the single trajectory can be defined in time and space by the series of subsequent temporary points 
\[\tau_1,r_1; \tau_2,r_2; \ldots ; \tau_n, r_n\]
 and the process can be described by the conditional probability density 
\[\rho(r_n,\tau_n \, | \, r_{n-1}, \tau_{n-1}; r_{n-2}, \tau_{n-2} ; \, \dots \, ;  r_2, \tau_2; r_1, \tau_1)\].
This is the probability density of jump increment $r_n$ after waiting time $\tau_n$, conditioned by the whole history $(\tau_1,r_1; \tau_2,r_2; \ldots ; \tau_{n-1}, r_{n-1})$. To construct theoretical model, we have to make the following simplifying assumptions: 
\begin{itemize}
\item the process is stationary, ergodic and homogeneous in space (price) variable. In the case of financial market, we neglect the influence of the so-called \emph{lunch effect}, which is the non-stationarity resulting as a daily stable pattern of investors' activity;
\item all waiting times between successive changes of the process, $\tau_n$, are i.i.d. random variables with distribution $\psi (\tau_n)$ having finite average\footnote{The stationary process we can obtain by using a modified distribution for the first jump, as we consider further in the text.}. In case of infinite average the process is non-ergodic \cite{bel2005,burov2010};
\item each jump increment $r_n$ of the process depends only on two previous jump increments $r_{n-1}, r_{n-2}$ in the form given by Eq. \refe{two:hh}.
\end{itemize}
The approximations given above can be summarized in the form of a factorized distribution, 
\bee
\rho&(&r_n,\tau_n \, | \, r_{n-1}, \tau_{n-1}; r_{n-2}, \tau_{n-2} ; \, \dots \, ;  r_2, \tau_2; r_1, \tau_1) \nonumber \\
&\approx &h(r_n \, | \, r_{n-1},r_{n-2}) \, \psi (\tau_n).
\label{rown:separable}
\eee
Equation \refe{rown:separable} gives the recipe for the infinitely long trajectory but, as the process is homogeneous and stationary, we can arbitrary choose the origin for the time and space axes. 
Since we analyze the trajectories starting at some arbitrary time $t=0$ at origin, we have to take into account that the first jump of the process after time $t=0$ depends on the two previous jumps, that we call $r_{0}$ and  $r_{-1}$. 
This can be solved by weighting the trajectories by $h( r_{0},r_{-1})$, where $h$ is given by Eq. \refe{eq:hold} even for $n=0$. 

Furthermore, we cannot use the same waiting-time distribution for the first jump as for other jumps. 
This is because jump increment $r_{0}$ might occur at any time before $t=0$. 
Therefore, we can average over all possible time intervals $\tau'$ between the instant of jump increment $r_{0}$ and the time origin $t=0$. 
Such an averaging was proposed in \cite{haus1987} and leads to the distribution
\bee
\psi_1 (\tau) &=& \frac{\ip{\tau'} \psi(\tau+\tau')}{\ip{\tau''}\ip{\tau'} \psi(\tau'+\tau'')} \nonumber \\
&\Leftrightarrow & 
\tilde{\psi}_1(s) = \frac{1}{\langle \tau \rangle }\frac{1-\psis}{s},
\label{psieq}
\eee
where expected (mean) waiting-time is  
\[\langle \tau \rangle = \int^{\infty}_{0} \tau \ \psi(\tau)d\tau < \infty \]. 
The denominator in the first equation in Eq. \refe{psieq} is required for the normalization. 
The only continuous case where $\psi_1 (\tau)=\psi(\tau)$ is an exponential waiting-time distribution of a Poisson process.  

The aim of this section is to derive the conditional probability density, $P\left(X, t \right)$, to find value $X$ of the process at time $t$, at condition that the process initial value was assumed as the origin. 
Further in the text we call this probability the \emph{soft} stochastic propagator, in contrast to the \emph{sharp} one, which we define below. 
The derivation of the propagator consists of few steps described in the following paragraphs, which extends the corresponding derivation of the canonical CTRW formalism.

The intermediate very useful quantity describing the stochastic process is the \emph{sharp}, $n$-step propagator 
\[Q_n \left( X, r_n, r_{n-1} ; t  \right), n=1,2,\ldots \] . 
This propagator is defined as the probability density that the process, which had initially (at $t=0$) the original value ($X=0$), makes its $(n-1)^{th}$ jump by $r_{n-1}$ from $X-r_n-r_{n-1}$ to $X-r_n$ (at any time) and makes its $n$-th jump by increment $r_n$ from $X-r_n$ to $X$ \underline{exactly} at time $t$. The key expression needed for exact solution of the process is given by the recursion relation 
\bee
& &\Q{n}{X, r_n, r_{n-1} ; t}=\int\limits_{0}^{t} d\tau \, \psi(\tau) \int\limits^{\infty}_{-\infty} d r_{n-2} \nonumber \\
&\times &\ h(r_n \mid r_{n-1}, r_{n-2}) \Q{n-1}{X -r_n, r_{n-1},r_{n-2};t-\tau}, \nonumber \\
n&=&3,4,
\label{rown:convbase}
\eee
Equation \refe{rown:convbase} relates two successive sharp propagators by the spatio-tempotral convolution. 
This equation is valid only for $n\geq 3$ and should be completed by propagators $Q_1$ and $Q_2$ calculated directly from their definitions (cf. Ref. \cite{gubiec2}).
 
We define \emph{sharp} summarized (infinite-many step) propagator $Q \left( X , t  \right)$ as follows,
\begin{eqnarray}
& &\Q{}{X,t}\stackrel{\rm def.}{=} \Q{1}{X ; t} + \Q{2}{X ; t} \nonumber \\
&+& \sum\limits_{n=3}^{\infty} \iall{r_n} \iall{r_{n-1}} \Q{n}{X, r_n, r_{n-1} ; t}
\label{rown:QSQn}.
\end{eqnarray}

Finally, to obtain the \emph{soft} stochastic propagator, $P \left( X , t  \right)$, we use the relation between \emph{soft} and \emph{sharp} propagators, which is much easier to consider in the Fourier-Laplace domain
\be
\hat{\tilde{P}}(k,s) = \tilde{\Psi}_1(s)+\tilde{\Psi}(s)\hat{\tilde{Q}}(k;s),
\label{rown:Pks}
\ee
where $\tilde{O}$ means the Laplace, and $\hat{O}$  Fourier transform of $O$. Sojourn probabilities (in time and Laplace domains) are defined by the corresponding waiting-time distribution
\be
\Psi(\tau) =\int\limits_{\tau}^{\infty} \psi(\tau') d\tau' \Leftrightarrow \tilde{\Psi }(s)=\frac{1-\tilde{\psi }(s)}{s},
\label{rown:Psit}
\ee
wherein $\Psi_1(\tau)$ is defined anologously.

To find an explicit form of Eq. \refe{rown:Pks} the procedure was developed analogous to that for the one-step memory model \cite{gubiec2}, although much more tedious 
(see Appendix \ref{section:AppendA} for details).
Hence, the Laplace transform of the velocity autocorrelation function (VAF) is given by
\bee
& &\tilde{C}(s)=\frac{s^2}{2} \left<\tilde{X}^2\right>(s)=\frac{\mu_2}{2 \tsr} \frac{N(\tpsi,\zeta,\epsilon)}{D(\tpsi,\zeta,\epsilon)},
\label{two:tC}
\eee
while numerator, $N(\tpsi,\zeta,\epsilon)$, and denominator, $D(\tpsi,\zeta,\epsilon)$, are defined as follows
\bee
N(\tpsi,\zeta,\epsilon)&=&\tpsi ^3 \zeta \left(\zeta - \epsilon ^2 \right)+\tpsi ^2 \epsilon \zeta (2 (\epsilon-1) \epsilon -\zeta +1) \nonumber \\
&+&\tpsi (1-\epsilon)^2 \left(2 \epsilon ^2-\zeta \right)-(1-\epsilon)^2 \epsilon, \nonumber \\
D(\tpsi,\zeta,\epsilon)&=&\tpsi ^3 \zeta \left(\zeta - \epsilon ^2 \right) + \tpsi ^2 \epsilon \zeta (\zeta -2 \epsilon +1)\nonumber \\
&+&\tpsi \zeta (1-\epsilon)^2    +(1-\epsilon)^2 \epsilon,
\label{rown:regularCC}
\eee
where the relation between the stochastic propagator in the Fourier and Laplace domains and the corresponding mean-square displacement was used herein.

As we are interested in a closed form of the VAF in time domain, we find both expressions in Eq. (\ref{rown:regularCC}) as too complicated to perform the inverse Laplace transformation of Eq. (\ref{two:tC}), even for simple forms of $\psis$. 
To keep our model self-consistent (that is, Eq. \refe{eq:h2} being an extension of Eq. \refe{eq:h2wrong}), we assume
\be
\zeta = \epsilon^2. \label{eq:assumption} 
\ee
Our estimation of parameter $\zeta $, based on the empirical data, gives this parameter almost equals $\epsilon^2$. Hence, relation \refe{eq:assumption} simplifies both expressions in Eq. \refe{rown:regularCC} eliminating the residual fluctuations of the order of $\zeta - \epsilon^2$.
Eq. \refe{two:tC} is simplified now into the more useful (formally) quite different forms,
\bee
\tilde{C}(s)&=&\frac{\mu_2}{2 \tsr} \frac{1-\epsilon \psis-\epsilon^2 \tpsi^2 (s)}{1+\epsilon \psis+\epsilon^2 \tpsi^2 (s)}=\frac{\mu_2}{2 \tsr}  \nonumber \\
&\times&\left[ 1-2 \left(\frac{j}{j-1} \ \frac{\epsilon \psis }{\epsilon \psis-j}+\frac{-1}{j-1} \ \frac{\epsilon \psis }{\epsilon \psis-\bar{j}} \right) \right], \nonumber \\
\label{rown:regularC}
\eee
where root $j=-\frac{1}{2}+ i \frac{\sqrt{3}}{2}$ and $\bar{O}$ means a complex conjugate of $O$.

It is worth  mentioning that we can obtain power spectra of our process from Eq. (\ref{rown:regularC}) directly by using Wiener-Khinchin theorem \cite{kubo}. 
The normalized VAF is given, in time domain, by expression
\bee
C^{\, n} (t) &=&\delta(t)-2 \epsilon \mathcal{L}^{-1}_{t} \left(\lambda \frac{ \psis }{\epsilon \psis-j} +
\bar{\lambda}\frac{ \psis }{\epsilon \psis-\bar{j}}\right), \nonumber \\
\label{two:C}
\eee
where ${\cal L}_t^{-1}\{\ldots \}$ is an inverse Laplace transform and  $\lambda=\frac{j}{j-1}=\frac{1}{2}- i \frac{\sqrt{3}}{6}$.
Apparently, the above obtained VAF uses solely the quantities ($\psis$ and parameter $\epsilon$) analogous to that of the one-step memory model \cite{gubiec2}.

Moreover, the very regular form of Eq. (\ref{rown:regularC}) and the corresponding result for the model containing the one-step memory backward enables to formulate the conjecture concerning the memory through arbitrary number of steps
\bee
\tilde{C}(s)&=&\frac{\mu_2}{2 \tsr} \frac{2-\sum_{j=0}^{n}\epsilon^j\psisj}{\sum_{j=0}^{n}\epsilon^j\psisj} \nonumber \\
&=&\frac{\mu_2}{2 \tsr}\frac{1-2\epsilon \psis + (\epsilon \psis )^{n+1}}{1 - (\epsilon \psis )^{n+1}},\; n=1,2,\ldots , \nonumber \\
\label{rown:regularCmany}
\eee
where three terms in denominator of the first equality in Eq. (\ref{rown:regularC}) are treated, herein, as initial terms of a geometric series (accordingly, the numerator is treated).
Apparently, for infinite many steps backward, $(n\rightarrow \infty)$, this equation gives the following result,
\bee
\tilde{C}(s)&=&\frac{\mu_2}{2 \tsr} \frac{2-\sum_{j=0}^{\infty}\epsilon^j\psisj}{\sum_{j=0}^{\infty}\epsilon^j\psisj}
=\frac{\mu_2}{2 \tsr}[1-2\epsilon \psis ]. \nonumber \\
\label{rown:regularCinf}
\eee
very useful for our further considerations. 

This is a significant issue that the evolution of $\tilde{C}(s)$ is govern in Eqs. (\ref{rown:regularCmany}) and (\ref{rown:regularCinf}) solely by $\psis $. For instance, a multifractality can be directly considered using Eq. (\ref{rown:regularCinf}) if $\psi (t)$ would be conducted in the form of properly suited superstatistics \cite{perello2008,kasprzak2010}. However, analysis of anomalous diffusion requires, herein, resignation from stationarity. Lack of stationarity, which would appear here, results from the initial situation and not with how the process itself evolves. Therefore, there are no major obstacles to build a non-stationary formalism CTRW containing memory through infinite-many steps. These concepts define research environment in which ergodicity  as well as the Bernoulli law of large numbers and hence the Wiener-Khintchine theorem would be broken while central limit theorem should be extened to the L\'evy-Khintchine theorem of the canonical representation of stable laws. However, this is already a subject of a separate work.

\section{Comparison of the models} \label{sec:com}

In Sec. \ref{sec:mot} we discussed selected properties of the one-step memory model and compared them with well known properties of empirical data.
Observed disagreement served as inspiration for development of the two- and infinite-step memory model solved in the previous section, where the latter model is based on our conjuction. In the presnt section we study the difference between one-, two-, and infinite-step memory models by using so much significant characteristic as autocorrelation function.

Let us begin with the analysis of the autocorrelation function in the event time.
The dependence between any two jumps of the process within the one-step memory model is given by Eq. \refe{eq:hkwrong}. 
This dependence results in autocorrelation expressed by Eq. \refe{eq:ckwrong}. 
However, in the case of the two-step memory model, the analogous dependence for $k \geq 1$ is a bit more complicated.
Fortunately, we obtain autocorrelation functions in the event time solely to $k=4$.
These functions are compared in Tab. \ref{tab:12} with the corresponding results for one-step memory model calculated from  Eq. \refe{eq:ckwrong}.
\begin{table}[ht]
	\caption{Comparison of the two-step memory model with the one-step memory model. Subsequent five values of the event-time autocorrelation function are presented. Apparently, the two-step memory model gives $c(2)=0$ as required by empirical data.}
	  \label{tab:12}
	\centering
		\begin{tabular}{l | c c c c c}
			                &$c(0)$& $c(1)$      & $c(2)$       & $c(3)$        & $c(4)$      \\ \hline
			one-step memory & $1$  & $-\epsilon$ & $\epsilon^2$ & $-\epsilon^3$ & $\epsilon^4$ \\
			two-step memory & $1$  & $-\epsilon$ & $0$          & $\epsilon^3$  & $-\epsilon^4$ \\
			empirical data & $1$ & $-\epsilon$ & $0.0$ & $0.0$ & $0.0$ \\
		\end{tabular}
\end{table}
It's worth recalling that empirical lack of the correlation for $k \geq 2$ is considered as a stylized fact in high frequency empirical financial data. 
As we assumed during the construction of the model, the current version gives $c(2)=0$.  
Apparently, for $k \geq 3$ both models give autocorrelation functions of the same orders, which for empirical values of $\epsilon$ are negligibly small quantities (see also Sec. \ref{sec:emp}).

The particularly useful way to visualize the role of the two- and infinite many step memory models is to compare the corresponding velocity autocorrelation functions in a real time. We perform the inverse Laplace transformation in Eqs. \refe{two:C} and \refe{rown:regularCinf} for the simplest possible exponential waiting-time distribution to highlight the generic difference between the models. 
We assume
\bee
\psit = \frac{1}{\tsr} e^{-t/\tsr}, \label{one:oneexp}
\eee
where $\tsr$ is an average of interevent time. 
Notably, this is solely waiting-time distribution obeying the equality $\psitf=\psit$ for the first step.
In the frame of one-step memory model this WTD leads to the normalized VAF in the form (cf. Eq. (23) in \cite{gubiec2})
\bee
C^{\, n}(t)=\delta (t) - \frac{2 \epsilon }{\tsr} e^{-(1+\epsilon)t/\tsr}.
\label{one:Coneexp}
\eee
Analogously, by substituting Eq. \refe{one:oneexp} into Eq. \refe{two:C} we obtain for the two-step memory model
\bee
C^{\, n}(t)
=\delta (t) - \frac{2 \epsilon }{\tsr}  e^{-(1+\epsilon/2) t/ \tsr} \frac{2}{\sqrt{3}}  \cos\left( \frac{\sqrt{3} \epsilon}{2} \frac{t}{\tsr} + \frac{\pi}{6}\right).
\nonumber \\
\label{one:C2oneexp}
\eee
It can be easily proved that for $\epsilon \ll 1$ the VAF given by Eq. \refe{one:C2oneexp} reduces into Eq. \refe{one:Coneexp}. 
This reduction was expected as within approximation given by Eq. \refe{eq:assumption} the difference between both models is of the order of $\epsilon^2$.

Furthermore, combining the conjecture given by Eq. (\ref{rown:regularCinf}) and the exponential WTD defined by Eq. (\ref{one:oneexp}) we easily obtain
\bee
C^{\, n}(t)=\delta (t) - \frac{2\epsilon }{\tsr} e^{-\epsilon t/\tsr}.
\label{rown:exactexp}
\eee

\begin{figure}[ht]
	\centering
		\includegraphics[width=0.47\textwidth]{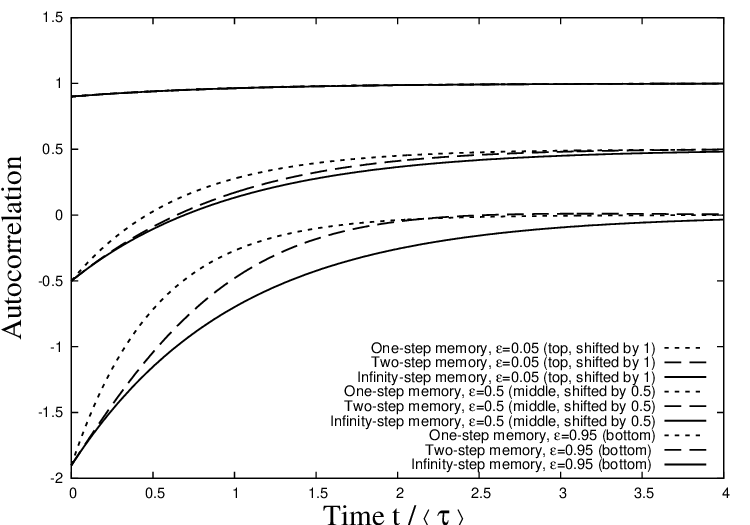}
	\caption{Comparison of velocity autocorrelation functions for one-, two-, and infinite many steps memory models for three different values of parameter $\epsilon=0.05, 0.5, 0.95$. All curves are based on exponential waiting time distribution given by Eq. \refe{one:oneexp}. For better visualization, the result for $\epsilon=0.05$ was shifted up by $1$, while result for $\epsilon=0.5$ up by $0.5$. For the same reason, all results are deprived of the Dirac's delta.}
	\label{fig:one_and_two_step}
\end{figure}

The predictions given by Eqs. \refe{one:Coneexp} -- \refe{rown:exactexp} for three different values of $\epsilon$ are shown in Fig. \ref{fig:one_and_two_step}.
As expected, for time
$t \gg \tsr$ the autocorrelations vanish within all three models, while initially autocorrelations begin their evolution from the same value.
For the intermediate time, more steps of memory taken into account in the model lead to a strengthen of the VAF. Besides, the increase of parameter $\epsilon$ increases the difference between autrocorrelation functions. 

\section{Comparison with empirical data} \label{sec:emp}

The satisfactory comparison of predicted autocorrelation functions with empirical ones requires \cite{gubiec2}:
\begin{itemize}
	\item[(i)] the use of sufficiently realistic waiting time distribution $\psis$, 
	\item[(ii)] determination of values of our basic parameter $\epsilon$ from the separate fit to the corresponding empirical histograms, and 
	\item[(iii)] the use of sufficiently effective method of VAF estimation for unevenly spaced elements of time-series (as interevent time intervals have random lengths).
\end{itemize}

Useful form of waiting-time distribution is a superposition (or weighted sum) of two exponential distributions \cite{gubiec2}
\bee
\psi (t)&=&\frac{w}{\tau _1} e^{-t/\tau _1}+\frac{1-w}{\tau _2} e^{-t/\tau _2} \nonumber \\
&\Leftrightarrow &\psis = \frac{w}{1+s \tau _1} + \frac{1-w}{1+s \tau _2},
\label{eq:WTD2exp}
\eee
where $0\leq w\leq 1$ is the weight, while $\tau _1$ and $\tau _2$ are the corresponding (partial) relaxation times. 
Apparently, this waiting time pdf has sufficiently simple (for the analytical calculations) closed form after the Laplace transformation. 
Such a form can be easily and satisfactory fitted to the empirical histogram of waiting times (cf. Fig. 4 in ref. \cite{gubiec2}).
Besides, this WTD makes the inverse Laplace transformation present in Eq. \refe{two:C} an analytically solvable. 
Finally, we obtain VAF in the useful form  
\bee
C^n(t)&=&\delta(t)-2 \epsilon \Big[\lambda \left( A_1 (j) \, e^{-\eta_1 (j) \, t} + A_2 (j) \, e^{-\eta_2 (j) \, t} \right) \nonumber \\
 &+& \bar{\lambda} \left( A_1 (\bar{j}) \, e^{-\eta_1 (\bar{j}) \, t} + A_2 (\bar{j}) \,  e^{-\eta_2 (\bar{j}) \, t} \right) \Big], 
 \label{two:Ctwoexp}
\eee
where
\bee
\eta _{1,2} (j) &=&\frac{1}{2 j}\left[j (\omega_1+\omega _2) + \epsilon\upsilon \pm \gamma_{1,2}(j)\right], \nonumber \\
\gamma_{1,2}(j)&=&\sqrt{\left(j (\omega _1+\omega _2)+\epsilon\upsilon \right)^2 - 4j (j+\epsilon) \omega _1\omega _2},\nonumber \\
A_1 (j) &=& \frac{\omega _1\omega _2-\eta _1 \upsilon}{j \; (\eta _2-\eta _1)},\nonumber \\
A_2 (j) &=& \frac{\omega _1\omega _2-\eta _2 \upsilon}{j \; (\eta _1-\eta _2)},\nonumber \\
\omega_1&=&1/\tau_1, \nonumber \\
\omega_2&=&1/\tau_2, \nonumber \\
\upsilon &=& w \; \omega _1 + (1-w)\; \omega _2 . \label{two:A12}
\eee
Apparently, above given VAF is a nontrivial because it contains the complex prefactors and exponents, which makes its interpretation a more complicated. 


Notably, all required parameters $\tau _1, \tau _2, w$ and $\epsilon$ are estimated by the separate empirical data without exploiting the empirical VAF. This is a basic result for further considerations.
That is, to find parameter $\epsilon$ only set of empirical jump increments are sufficient to have, while for the estimation of remaining parameters only set of empirical waiting times is required and available. We obtained a very promising comparison of our theoretical VAF with its empirical counterpart because this is not a fit as no free parameters were left to make it.

\begin{figure}[ht]
\includegraphics[width=0.47\textwidth]{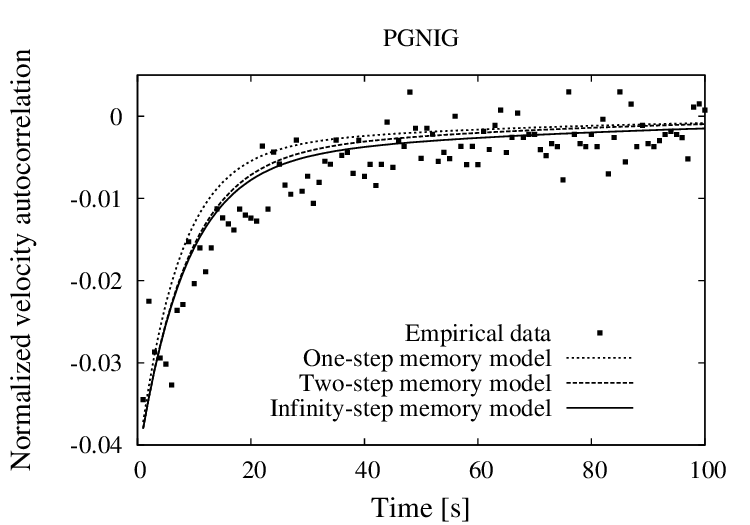}
\caption{Comparison of three normalized autocorrelation functions of price velocity, for instance, for the PGNIG company (from the fuel and energy sector) quoted on the Warsaw Stock Exchange: (i) the empirical one (small black squares), (ii) the theoretical VAF (dotted curve) derived in Ref. \cite{gubiec2} for the one-step memory model, (iii) the theoretical VAF (dashed curve) for the two-step memory model, and (iv) the theoretical VAF (solid curve) for the infinite many-step memory model. All models use the waiting time distribution in the form of the weighted sum of two exponential distributions given by Eq. \refe{eq:WTD2exp}. All parameters were fitted to separate data records, so both theoretical curves have no free parameters to fit to the empirical VAF (notably, the required method of determination of the empirical VAF was thoroughly considered in ref. \cite{gubiec2}).}
\label{two:f:autocorr}
\end{figure}

The comparison of our theoretical predictions with the corresponding empirical VAFs is shown in Fig. \ref{two:f:autocorr}.
The improved agreement provided by the two- and infinite many step memory models in comparison with the one-step memory model is well seen. However, still small but distinct systematic deviations exist even for the infinite many step memory model (especially for the initial time).

\section{Concluding remarks}\label{sec:sum}

Hithertoo, only one step backward memory was considered analytically. In the present paper we developed the version of the CTRW formalism which contains memory over two steps backward or dependence between three consecutive jumps of the process. This extended dependence was studied, herein, independently on whether the second order correlations in the system exist or not, which significantly extends the capability of the CTRW formalism. Such an approach suggests that several already existing results could be improved if one step backward memory model used there, would be changed by two step backward memory model. 


There are two results which can be considered as an achievement of our work:
\begin{itemize}
\item[(i)] the derivation of the analitycal closed formula for the propagator containing the two-step memory (cf. Eq. \refe{rown:Pks}), which is a significant extension of the corresponding one from our previous paper (cf. Eq. (17) in Ref. \cite{gubiec2}), without increasing the number of free parameters and functions,
\item[(ii)] the conjugtion which enabled derivation of the velocity autocorrelation function containing infinite many step memory, keeping the model still analytically solvable -- the solution is even simpler than for those with finite-step memory.
\end{itemize}

By using propagator \refe{rown:Pks} we obtained the velocity autocorrelation of the process which was compared with its empirical counterpart (analogously as it was performed in \cite{gubiec2}, cf. plot in Fig. \ref{two:f:autocorr} in the present paper) strongly improving the agreement.



The approach presented in this work brings a new approach to random walks with memory, because shows that even in the case of a finite diffusion coefficient the dependence through infinitely many steps play an important role in the CTRW formalism, making the VAF much more realistic. However, still presence of a distinct deficit of correlations suggest that perhaps dependence between interevent times should be somehow coupled with a multi-step memory -- this is still a challenge.   

\appendix

\section{Proof}\label{section:Proof}

Below we prove that $c(2)$ is always larger than 0 if only two successive jumps are dependent.
\bee
c(2)&=&\frac{1}{\mu_2}\iall{r_{n}}\iall{r_{n-1}} \iall{r_{n-2}} \ r_{n} \ r_{n-2} \nonumber \\
&\times & h (r_n \mid r_{n-1}) \ h (r_{n-1},r_{n-2}) \nonumber \\
&=&\frac{1}{\mu_2}\iall{r_{n-1}} \ h (r_{n-1})  \iall{r_{n}} \ r_{n} \  h (r_n \mid r_{n-1}) \nonumber \\
&\times &\iall{r_{n-2}} \ r_{n-2} \ h (r_{n-2} \mid r_{n-1}) \nonumber\\
&=&\frac{1}{\mu_2}\iall{r_{n-1}} \ h (r_{n-1}) \nonumber \\
&\times &\left(\iall{r_{n}} \ r_{n} \  h (r_n \mid r_{n-1}) \right)^2 > 0.
\eee

\section{Detailed derivation}\label{section:AppendA}

\renewcommand{\refe}[1]{(\ref{#1})}
\renewcommand{\AAA}{\left(\frac{1-2 \epsilon+\theta}{1-\epsilon}\right)}
\renewcommand{\BB}{\left(\frac{\zeta-\theta}{1-\epsilon}\right)}
\renewcommand{\CC}{\left(\frac{\epsilon-\zeta}{1-\epsilon}\right)}
\renewcommand{\DD}{\left(\frac{\epsilon-\zeta}{\epsilon}\right)}
\renewcommand{\EE}{\left(\frac{\zeta}{\epsilon}\right)}
In this appendix we present detailed procedure of obtaining the propagator of out 2-step memory CTRW model. Let us start with the generalization of the pdf of three consecutive jumps $h(r_3,r_2,r_1)$ given by Eq. \refe{eq:h321}. Our 2-step memory model is coherent with the 1-step memory model presented in \cite{gubiec2} only at the level of two-variable pdfs $h(r_3,r_2)$ and $h(r_2,r_1)$. Dependance $h_2(r_3,r_1)$ present in 2-step memory model give by \refe{eq:h2} is irreducible to the form obtained within 1-step memory model \refe{eq:h2wrong}. To obtain such a reducibility we propose to add another, formal parameter $\theta$ in  $h(r_3,r_2,r_1)$ form:
\bee 
h(r_3,r_2,r_1) &=& (1-2\epsilon+\theta) h(r_1)h(r_2)h(r_3) \nonumber \\
&+&\zeta \delta(r_3+r_2)\delta(r_2+r_1)h(r_1) \nonumber \\
&+& (\zeta-\theta) \delta(r_3+r_1)h(r_2)h(r_1) \nonumber \\
&+& (\epsilon -\zeta) \delta(r_2+r_1)h(r_3)h(r_1) \nonumber \\
&+& (\epsilon -\zeta) \delta(r_3+r_2)h(r_2)h(r_1). 
\label{two:h321} 
\eee
As a result, for $\theta=0$ Eq. (\ref{two:h321}) reduces to the form of \refe{eq:h321} and for $\theta=\zeta=\epsilon^2$ we obtain 1-step memory model.


For simpler notation let us introduce variable:
\bee
A&=&\AAA, \\
B&=&\BB, \\
C&=&\CC, \\
D&=&\DD, \\
E&=&\EE.
\eee
Conditional pdf $h(r_3 \mid r_2,r_1)$ can be now expressed in the form
\bee
h(r_3 \mid r_2, r_1) &=& \left(1-\delta_{r_2,-r_1}\right) \Big[A h(r_3) + B \delta(r_3+r_1) \nonumber \\
&+& C \delta(r_3+r_2)\Big] + \nonumber \\
&+& \delta_{r_2,-r_1}\left[D h(r_3) + E \delta(r_3+r_2) \right]. \label{two:h}
\eee

The key relation needed for exact solution of the propagator is given by Eq. \refe{rown:convbase} which we recall below
\bee
\Q{n}{X, r_n, r_{n-1} ; t }&=&\int\limits_{0}^{t}dt^{\prime }\psi(t^{\prime }) \nonumber \\
&\times &\int\limits^{\infty}_{-\infty} d r_{n-2} 
h(r_n \mid r_{n-1}, r_{n-2}) \nonumber \\
&\times &\Q{n-1}{X -r_n, r_{n-1},r_{n-2};t-t^{\prime } }, \nonumber \\
\label{A:rekur}
\eee
For the simplicity of notation let us introduce a notation
\bee
\tQ{n}{r_n, r_{n-1}} &\equiv &\int\limits_{0}^{\infty} dt \, \int\limits^{\infty}_{-\infty} dX \edo{i k X} \edo{- s t} \nonumber \\
&\times &\Q{n}{X, r_n, r_{n-1} ; t}, 
\label{A:Qzapis}
\eee
where we omit an explicit depandance on varaibles $k$ and $s$.

The key relation \refe{A:rekur} in Fourier--Laplace space and with notation defined above, takes the form:
\bee
 \tQ{n}{r_n,r_{n-1}}&=&\psis \edo{i k r_n} \iall{r_{n-2}}  \ h(r_n \mid r_{n-1}, r_{n-2}) \nonumber \\
&\times &\tQ{n-1}{r_{n-1}, r_{n-2}}.
\eee
As previously, to obtain more intuitive notation we change variables  ($r_n, r_{n-1},r_{n-2}$) to ($r_3,r_2,r_1$), what gives
\bee
\frac{1}{\psis} \tQ{n}{r_3,r_2}&=& \edo{i k r_3} \iall{r_1}  \ h(r_3 \mid r_2, r_1) \nonumber \\
&\times &\tQ{n-1}{r_2, r_1}.
\label{A:trekur}
\eee
The method which will allow us to solve equation \refe{A:trekur}, for the given form of  $\ h(r_3 \mid r_2, r_1)$ in \refe{two:h}, is to seperate sharp propagator  $\tQ{n}{r_3,r_2}$ into two parts: singular and regular given accordingly by :
\bee
\tQR{n}{r_2,r_1} = \left(1-\delta_{r_2,-r_1}\right) \tQ{n}{r_2,r_1}, \label{A:QR}\\
\tQS{n}{r_2} =\iall{r_1} \delta_{r_2,-r_1} \tQ{n}{r_2,r_1}. \label{A:QS}
\eee
With the quantities above we can restore $\tQ{n}{r_3,r_2}$, using relation
\bee
\tQ{n}{r_2,r_1} = \tQR{n}{r_2,r_1} + \delta(r_2+r_1) \tQS{n}{r_2}. \label{A:QQRQS}
\eee
Next, we transform relation \refe{A:trekur} substituting $h$ with its exact form \refe{two:h} and using definitions \refe{A:QR} i \refe{A:QS}:
\bee
& &\edo{- i k r_3} \frac{\tQ{n}{r_3,r_2} }{\psis} = \tQS{n-1}{r_2} \left[D h(r_3) + E \delta(r_3+r_2) \right] \nonumber \\
&+&B \tQR{n-1}{r_2, -r_3} \nonumber \\
&+& \iall{r_1}  \tQR{n-1}{r_2, r_1} \left[A h(r_3) + C \delta(r_3+r_2) \right]. 
\label{A:rekur1}
\eee
The RHS of \refe{A:rekur1} contains only regular and singular propagators, while on the LHS we still have full propagator $\tQ{n}{r_3,r_2}$. 
Our aim is to obtain recurance relation between regular and singular propagator.

Let us multiply both sides of the equation above by $\left(1-\delta_{r_3,-r_2}\right)$
\bee
& &\edo{- i k r_3} \left(1-\delta_{r_3,-r_2}\right) \frac{\tQ{n}{r_3,r_2}}{\psis}=\tQS{n-1}{r_2} \nonumber \\
&\times &\left(1-\delta_{r_3,-r_2}\right) D h(r_3) \nonumber \\
&+& B \left(1-\delta_{r_3,-r_2}\right) \tQR{n-1}{r_2, -r_3}+ \left(1-\delta_{r_3,-r_2}\right) \nonumber \\
&\times &A h(r_3) \iall{r_1}  \tQR{n-1}{r_2, r_1}, 
\label{A:rekur1R}
\eee
as a result the term containing $\left(1-\delta_{r_3,-r_2}\right) \delta (r_3+r_2)$ disappear.
Afterwards, we multiply both sides by $\delta_{r_3,-r_2}$, what leads to
\bee
& &\edo{- i k r_3} \frac{1}{\psis} \delta_{r_3,-r_2} \tQ{n}{r_3,r_2} = \tQS{n-1}{r_2} \nonumber \\
&\times &\left[D \delta_{r_3,-r_2} h(r_3) + E \delta(r_3+r_2) \right] 
+B \delta_{r_3,-r_2} \tQR{n-1}{r_2, -r_3} \nonumber \\
&+&\iall{r_1} \tQR{n-1}{r_2, r_1} \left[A \delta_{r_3,-r_2} h(r_3) + C \delta(r_3+r_2) \right]. \nonumber \\
\label{A:rekur1S}
\eee

Now we rewrite relations \refe{A:rekur1R} and \refe{A:rekur1S} in terms of new variables integrating the latter over $r_2$\footnote{We neglect zero measure sets and assume that pdf $h(x)$ does not contain any term proportional to Diraca's delta}
\bee
\edo{- i k r_3} \frac{\tQR{n}{r_3,r_2}}{\psis}&=& 
D h(r_3) \tQS{n-1}{r_2} + B \tQR{n-1}{r_2, -r_3} \nonumber \\
&+&A h(r_3) \iall{r_1}  \tQR{n-1}{r_2, r_1}, 
\label{A:rekur2R}
\eee
\be
\edo{- i k r_3} \frac{\tQS{n}{r_3}}{\psis}=E \tQS{n-1}{-r_3} 
+ C \iall{r_1}  \tQR{n-1}{-r_3, r_1}. 
\label{A:rekur2S}
\ee
As a result we obtained system of two recurrence equations on regular and singular sharp propagators.
In order to solve the equations above let us introduce additional functions
\bee
\textbf{R}_n (b,a) &=& \Re \iall{r_3} \edo{i b k r_3} \iall{r_2} \edo{i a k r_2} \tQR{n}{r_3,r_2}, \nonumber \\
\label{A:defR}\\
\textbf{S}_n (a) &=& \Re \iall{r} \edo{i a k r} \tQS{n}{r}, 
\label{A:defS}\\
H (a) &=& \Re \iall{x} \edo{i a k x} h(x), 
\label{A:defH}
\eee
where operator $\Re$ giver the real part of the complex number.
Due to the definition the following properties are satisfied
\bee 
\textbf{R}_n (b,a)&=&\textbf{R}_n (-b,a)=\textbf{R}_n (b,-a)=\textbf{R}_n (-b,-a), \nonumber \\
\textbf{S}_n (a)&=&\textbf{S}_n (-a), \nonumber \\
H (a)&=&H (-a),
\label{A:wlasnosci} 
\eee
or simply the functions are even in all their parameters.

Acting on both sides of Eq. \refe{A:rekur2R} with the operator 
\be
\Re \iall{r_3} \edo{i b k r_3} \iall{r_2} \edo{i a k r_2}, \nonumber
\ee
on Eq. \refe{A:rekur2S} with operator 
\be
\Re \iall{r_3} \edo{i a k r_3} \nonumber
\ee
and using definitions \refe{A:defR} i \refe{A:defS} we obtain
\bee
\frac{\textbf{R}_n (b-1,a)}{\psis}&=&D \ H(b) \ \textbf{S}_{n-1}(a) + B \ \textbf{R}_{n-1} (a,-b) \nonumber \\
&+& A \ H(b) \ \textbf{R}_{n-1} (a,0), \label{A:rekurR} \nonumber \\
\frac{\textbf{S}_n (b-1)}{\psis} &=& E \ \textbf{S}_{n-1} (-b)+ C \ \textbf{R}_{n-1} (-b, 0). \label{A:rekurS}
\eee

Notably, by basing on \refe{rown:QSQn} in Fourier--Laplace domain, Eq. \refe{A:QQRQS} and definitions \refe{A:defR}, \refe{A:defS} we obtain
\bee 
\tQ{n}{k, s }&=&\iall{r_2} \iall{r_1} \ \tQ{n}{k, r_2, r_1 ; s} \nonumber \\
&=&\iall{r_2} \iall{r_1} \ \tQR{n}{r_2,r_1} + \iall{r_2} \ \tQS{n}{r_2} \nonumber \\
&=&\textbf{R}_n (0,0) + \textbf{S}_n (0).
\label{A:QnRnSn}
\eee

Apparently, the sharp propagator $\tQ{n}{k, s }$ can be expressed by functions $\textbf{R}_n$ and $\textbf{S}_n$  with all arguments equal to zero.
Hence, our aim is to obtain $\textbf{R}_n$ and $\textbf{S}_n$  with all arguments equal to zero from the recurrance relationns \refe{A:rekurR} and \refe{A:rekurS}.
Let us start with analysis of \refe{A:rekurR} in four cases, using properties \refe{A:wlasnosci}:

\begin{eqnarray*}
\left(
\begin{array}{c}
a=0 \\ b=0 
\end{array}
\right) \, \Rightarrow
\frac{\textbf{R}_n (1,0)}{\psis} &=& D \ H(0) \ \textbf{S}_{n-1}(0)+B \ \textbf{R}_{n-1} (0,0) \nonumber \\
&+&A \ H(0) \ \textbf{R}_{n-1} (0,0), 
\end{eqnarray*}
\begin{eqnarray*}
\left(
\begin{array}{c}
a=0 \\ b=1 
\end{array}
\right) \, \Rightarrow 
\frac{\textbf{R}_n (0,0)}{\psis} &=& D \ H(1) \ \textbf{S}_{n-1}(0)+B \ \textbf{R}_{n-1} (0,1) \nonumber \\
&+&A \ H(1) \ \textbf{R}_{n-1} (0,0),
\end{eqnarray*}
\begin{eqnarray*}
\left(
\begin{array}{c}
a=1 \\ b=0 
\end{array} 
\right) \, \Rightarrow
\frac{\textbf{R}_n (1,1)}{\psis}&=&D \ H(0) \ \textbf{S}_{n-1}(1) + B \ \textbf{R}_{n-1} (1,0) \nonumber \\
&+&A \ H(0) \ \textbf{R}_{n-1} (1,0), 
\end{eqnarray*}
\begin{eqnarray}
\left(
\begin{array}{c}
a=1 \\ b=1 
\end{array} 
\right) \, \Rightarrow
\frac{\textbf{R}_n (0,1)}{\psis}&=&D \ H(1) \ \textbf{S}_{n-1}(1) + B \ \textbf{R}_{n-1} (1,1) \nonumber \\
&+& A \ H(1) \ \textbf{R}_{n-1} (1,0). 
\label{A:matrix}
\end{eqnarray}
At this place it is resonalble to redefine factors $A,B,C,D,E$ to remove $\psis$ from the relation, what gives 
\bee
A=&\psis \AAA, \nonumber \\
B=&\psis  \BB, \nonumber \\
C=&\psis \CC, \nonumber \\
D=&\psis \DD, \nonumber \\
E=&\psis \EE.
\label{two:ABCDE}
\eee
After few linear operations on the system of equations \refe{A:matrix} we obtain an expression
\begin{eqnarray*}
B^3 \textbf{R}_{n-3} (1,0)&=& B^3 D \ H(0) \ \textbf{S}_{n-4}(0) + B^4 \ \textbf{R}_{n-4} (0,0) \nonumber \\
&+&B^3 A \ H(0) \ \textbf{R}_{n-4} (0,0), 
\end{eqnarray*}
\begin{eqnarray*}
\textbf{R}_n (0,0)&=&D \ H(1) \ \textbf{S}_{n-1}(0) + B \ \textbf{R}_{n-1} (0,1) \nonumber \\
&+&A \ H(1) \ \textbf{R}_{n-1} (0,0), 
\end{eqnarray*}
\begin{eqnarray*}
B^2 \textbf{R}_{n-2} (1,1)&=&B^2 D \ H(0) \ \textbf{S}_{n-3}(1) + B^3 \ \textbf{R}_{n-3} (1,0) \nonumber \\
&+&B^2 A \ H(0) \ \textbf{R}_{n-3} (1,0), 
\end{eqnarray*}
\begin{eqnarray*}
B \textbf{R}_{n-1} (0,1)&=&B D \ H(1) \ \textbf{S}_{n-2}(1) + B^2 \ \textbf{R}_{n-2} (1,1) \nonumber \\
&+&B A \ H(1) \ \textbf{R}_{n-2} (1,0). 
\end{eqnarray*}
Summation of both sides of given above equations gives
\begin{eqnarray}
\textbf{R}_n (0,0)& =& B^3 D \ H(0) \ \textbf{S}_{n-4}(0) + B^4 \ \textbf{R}_{n-4} (0,0) \nonumber \\
&+& B^3 A \ H(0) \ \textbf{R}_{n-4} (0,0) + D \ H(1) \ \textbf{S}_{n-1}(0) \nonumber \\
&+&  A \ H(1) \ \textbf{R}_{n-1} (0,0) + B^2 D \ H(0) \ \textbf{S}_{n-3}(1) \nonumber \\
& +&   B^2 A \ H(0) \ \textbf{R}_{n-3} (1,0) + B D \ H(1) \ \textbf{S}_{n-2}(1) \nonumber \\
&+&  B A \ H(1) \ \textbf{R}_{n-2} (1,0). 
\label{A:1R00}
\end{eqnarray}
On the RHS of  \refe{A:1R00} two components signed by $\textbf{R}_{n-3} (1,0)$ and $\textbf{R}_{n-2} (1,0)$ are still present. They can be once more expressed (with help of \refe{A:matrix}) by using
\begin{eqnarray*}
B^2 A\ H(0) \textbf{R}_{n-3} (1,0) &=&A B^2  D \ H^2(0) \ \textbf{S}_{n-4}(0) \nonumber \\
&+&A B^3  \ H(0) \textbf{R}_{n-4} (0,0) \nonumber \\
&+& A^2 B^2  \ H^2(0) \ \textbf{R}_{n-4} (0,0), 
\end{eqnarray*}
and
\begin{eqnarray*}
B A\ H(1) \textbf{R}_{n-2} (1,0) &=&  A B D \ H(0)\ H(1) \ \textbf{S}_{n-3}(0) \nonumber \\
&+& A B^2  \ H(1) \textbf{R}_{n-3} (0,0) \nonumber \\
&+&  A^2 B \ H(0) \ H(1) \ \textbf{R}_{n-3} (0,0).
\end{eqnarray*}
Using relation $H(0)=1$ and substituting above two expressions to Eq. \refe{A:1R00} we obtain
\begin{eqnarray}
\textbf{R}_n (0,0)&=&   A \ H(1) \ \textbf{R}_{n-1} (0,0) +  B^2 (A+B)^2 \ \textbf{R}_{n-4} (0,0) \nonumber \\
&+& A B (A+B) \ H(1) \textbf{R}_{n-3} (0,0) + A  B^2 D  \ \textbf{S}_{n-4}(0) \nonumber \\
&+& A B D \ H(1) \ \textbf{S}_{n-3}(0)+ D B^3 \ \textbf{S}_{n-4}(0) +  \nonumber \\
&+& D \ H(1) \ \textbf{S}_{n-1}(0)+ D B^2 \ \textbf{S}_{n-3}(1) \nonumber \\
&+& D B \ H(1) \ \textbf{S}_{n-2}(1).  
\label{A:3R00}
\end{eqnarray}
where function $\textbf{R}$ occurs with argument equal to zero. We still have the function $\textbf{S}$ with the argument equal one. We can express it differently using relation \refe{A:rekurS} for $b=0$, what gives
\bee
\textbf{S}_n (1)   &=& E \ \textbf{S}_{n-1} (0)+ C \ \textbf{R}_{n-1} (0, 0).  \label{A:Sn1}
\eee
Hence
\begin{eqnarray*}
D B^2 \textbf{S}_{n-3} (1)   &=& D B^2 E \ \textbf{S}_{n-4} (0) \nonumber \\
&+& D B^2 C \ \textbf{R}_{n-4} (0, 0), \nonumber \\
D B \ H(1) \ \textbf{S}_{n-2} (1)   &=& D B E\ H(1)  \ \textbf{S}_{n-3} (0)\nonumber \\
&+& D B C \ H(1)  \ \textbf{R}_{n-3} (0, 0),  
\end{eqnarray*}
what substituted to  Eq. \refe{A:3R00} gives
\begin{eqnarray}
\textbf{R}_n (0,0)&=&  A \ H(1) \ \textbf{R}_{n-1} (0,0) + D \ H(1) \ \textbf{S}_{n-1}(0) \nonumber \\
 &+&\left[B^2 (A+B)^2 + D B^2 C \right] \textbf{R}_{n-4} (0,0) \nonumber \\
&+& \left[A B (A+B) + D B C\right] \ H(1) \textbf{R}_{n-3} (0,0) \nonumber \\
&+& D B^2 \left(A + B + E\right) \textbf{S}_{n-4}(0) \nonumber \\
&+& B D \left( A  +  E \right) H(1) \ \textbf{S}_{n-3}(0) ,
\label{A:finRrekur}
\end{eqnarray}
which is the final form with functions $\textbf{R}$ i $\textbf{S}$ with arguments equal zero

To obtain second recurrence relation we rewrite \refe{A:rekurS} to the form
\begin{eqnarray*}
\textbf{S}_n (b-1)   &=& E \ \textbf{S}_{n-1} (-b)+ C \ \textbf{R}_{n-1} (-b, 0).    
\end{eqnarray*}
For $b=1$ it gives
\begin{eqnarray*}
\textbf{S}_n (0)   &=& E \ \textbf{S}_{n-1} (1)+ C \ \textbf{R}_{n-1} (1, 0) ,   
\end{eqnarray*}
where the term containing $\textbf{S}_{n-1} (1)$ is obtained from \refe{A:Sn1}. It lead to 
\begin{eqnarray*}
\textbf{S}_n (0)   &=& E \ \left( E \ \textbf{S}_{n-2} (0)+ C \ \textbf{R}_{n-2} (0, 0) \right)+ C \ \textbf{R}_{n-1} (1, 0).    
\end{eqnarray*}
The term containing $\textbf{R}_{n-1} (1, 0)$, as previously can be expressed from \refe{A:matrix}
\begin{eqnarray*}
\textbf{R}_{n-1} (1,0) = D \ \textbf{S}_{n-2}(0) + (B + A) \ \textbf{R}_{n-2} (0,0),
\end{eqnarray*}
what gives the final form of the second recurrence relation
\be
\textbf{S}_n (0) = (E^2 + C D) \textbf{S}_{n-2} (0) + C (A + B + E) \ \textbf{R}_{n-2} (0,0). \label{A:finSrekur}
\ee

Now we introduce additional quantities to simplify the notation
\bee
\textbf{S}_n &\equiv& \textbf{S}_n (0) , \label{A:SnSn0}\\
\textbf{R}_n &\equiv& \textbf{R}_n (0,0), \label{A:RnRn0} \\
\textbf{S} &=& \sum\limits_{n=1}^{\infty} \textbf{S}_n, \label{A:SSn} \\
\textbf{R} &=& \sum\limits_{n=1}^{\infty} \textbf{R}_n. \label{A:RRn}
\eee
Now we perform summation of Eq. \refe{A:finSrekur} from $n=3$ to $\infty$ and a similar summation of Eq. \refe{A:finRrekur} from $n=5$ to $\infty$
\begin{eqnarray*}
\left(\textbf{S} - \textbf{S}_1 - \textbf{S}_2 \right) &=& (E^2 + C D) \textbf{S} \nonumber \\
&+& C (A + B + E) \ \textbf{R}, 
\end{eqnarray*}
\begin{eqnarray*}
& &\left(\textbf{R} - \textbf{R}_1 - \textbf{R}_2 - \textbf{R}_3 - \textbf{R}_4 \right) \nonumber \\
&=&  A \ H(1) \ \left(\textbf{R} - \textbf{R}_1 - \textbf{R}_2 - \textbf{R}_3 \right) \nonumber \\
&+& D \ H(1) \ \left(\textbf{S} - \textbf{S}_1 - \textbf{S}_2 - \textbf{S}_3 \right) \nonumber \\
&+& \left[B^2 (A+B)^2 + D B^2 C \right] \textbf{R} \nonumber \\
&+& \left[A B (A+B) + D B C\right] \ H(1) \left(\textbf{R} - \textbf{R}_1 \right) \nonumber \\
&+& D B^2 \left(A + B + E\right) \textbf{S} + B D \left( A  +  E \right) H(1) \  \left(\textbf{S} - \textbf{S}_1 \right).
\end{eqnarray*}
By using $\tilh = \hk =H(1)$ we find the above given expressions as equivalent to
\begin{eqnarray}
0&=& \textbf{S}_1 + \textbf{S}_2 + (E^2 + C D -1 ) \textbf{S} + C (A + B + E)  \textbf{R}, \nonumber \\
0&=&\textbf{R} \left[-1 +A \ \tilh + B^2 (A+B)^2 + D B^2 C\right] \nonumber \\
&+&\textbf{R}\left[\left(A B (A+B) +D B C\right) \ \tilh \right] \nonumber \\
&+& \textbf{S} \left[D \ \tilh + D B^2 \left(A + B + E\right) +  B D \left( A  +  E \right) \tilh \right] \nonumber \\
&+& \textbf{R}_1 + \textbf{R}_2 + \textbf{R}_3 + \textbf{R}_4 \nonumber \\
&+& A  \ \left( - \textbf{R}_1 - \textbf{R}_2 - \textbf{R}_3 \right)\tilh  \nonumber \\
&+& \left[A B (A+B) + D B C\right]  \left(- \textbf{R}_1 \right) \tilh \nonumber \\
&+& D \ \left( - \textbf{S}_1 - \textbf{S}_2 - \textbf{S}_3 \right) \tilh  + B D \left( A  +  E \right) \left( - \textbf{S}_1 \right) \tilh . 
\label{A:uklad}
\end{eqnarray}
If we obtain explicitly the forms of  $\textbf{R}_1, \textbf{R}_2, \textbf{R}_3, \textbf{R}_4 , \textbf{S}_1 , \textbf{S}_2 , \textbf{S}_3$ we are able to reduce the problem of finding the sharp propagator to the problem of solving system of two equations \refe{A:uklad} with two variables $\textbf{R}$ i $\textbf{S}$.

In order to obtain explicit forms of  $\textbf{R}_1, \textbf{R}_2, \textbf{R}_3, \textbf{R}_4 , \textbf{S}_1 , \textbf{S}_2 , \textbf{S}_3$ we start with explicit form of first four propagators $Q$ calculated directly from the definition and presented in Fourier-Laplace space
\begin{eqnarray*}
\frac{\tQ{1}{k,s}}{\psisf}&=& \iall{\xi_0} \iall{X} \eiw{X} h (X,\xi_0), \nonumber \\
\frac{\tQ{2}{k,s}}{\psisf \psis}&=& \iall{r_1} \iall{X} \eiw{X} \eiw{r_1} \ h (X, r_1),  \nonumber \\ 
\frac{\tQ{3}{k,s}}{\psisf \tpsi^2 (s)}&=& \iall{r_1} \iall{r_2} \iall{X} \eiw{X}  \nonumber \\
& &\eiw{r_1}  \eiw{r_2} h(X, r_2, r_1),  \nonumber \\
\frac{\tQ{4}{k,s}}{\psisf \tpsi^3 (s)}&=& \iall{r_1} \iall{r_2} \iall{r_3} \iall{X} \eiw{X} \nonumber \\
& &\eiw{r_1}  \eiw{r_2}  \eiw{r_3} \ h(X \mid r_3, r_2) h(r_3, r_2, r_1).
\end{eqnarray*}
On the basis of Eqs. \refe{A:SnSn0} and \refe{A:RnRn0}, \refe{A:defR} and \refe{A:defS} as well as \refe{A:QR} and \refe{A:QS} with help of Eq. \refe{A:Qzapis} we obtain 
\begin{eqnarray*}
\frac{\textbf{S}_1}{\psisf}&=& \epsilon \tilh,\\
\frac{\textbf{R}_1}{\psisf}&=& (1-\epsilon) \tilh, \\
\frac{\textbf{S}_2}{\psisf \psis}&=& \epsilon,\\
\frac{\textbf{R}_2}{\psisf \psis}&=&(1-\epsilon) \tilh^2, \\
\frac{\textbf{S}_3}{\psisf \tpsi^2 (s)}&=& \epsilon \tilh, \\
\frac{\textbf{R}_3}{\psisf \tpsi^2 (s)}&=& (1-2\epsilon+\theta) \tilh^3 + (\epsilon-\theta) \tilh, \\
\frac{\textbf{R}_4}{\psisf \tpsi^3 (s)}&=& A (1-2\epsilon+\theta) \tilh^4 +B (\zeta-\theta) \\
&+& \quad \tilh^2 \left[ B (1-2\epsilon+\theta)  + D \zeta + A (\zeta-\theta)\right] \\ 
&+& \quad \tilh^2(A+B+D) (\epsilon -\zeta).
\end{eqnarray*}
Substituiting the above given equations into  \refe{A:uklad} and basing on  
\refe{rown:QSQn}, \refe{A:QnRnSn}, \refe{A:SSn} and \refe{A:RRn} we get
\begin{eqnarray}
\tilde{Q}(k, s)=\textbf{R}+\textbf{S} = \tpsi_1 \frac{N_1}{D_1} 
\label{A:Q}
\end{eqnarray}
where
\begin{eqnarray*}
N_1 &=& -(\epsilon -1) \tpsi ^3 (\zeta -\theta ) \nonumber \\ 
&\times & \left[\tilh^2 \left(\epsilon ^2-\zeta \right) \left(\epsilon ^2 (\zeta -\theta +1)-\zeta \right)-(\epsilon -1)^2 \epsilon ^2 (\zeta -\theta )\right] \nonumber \\
&+&(\epsilon -1)^3 \epsilon ^2 \tpsi  \left[\left(\tilh^2-1\right) \epsilon ^2+\tilh^2 (-\theta )+\epsilon \right] \nonumber \\
&-&\tilh (\epsilon -1)^3 \tpsi ^2 \left(\epsilon ^2 (\theta -2 \zeta )+\zeta ^2\right) \nonumber \\
&-& \tilh (\epsilon -1)^4 \epsilon ^2-\tilh \tpsi ^4 \left(\epsilon ^2-\zeta \right)(\zeta -\theta ) \nonumber \\
&\times &\left[\epsilon ^3+\epsilon ^2 (-\zeta +\theta -1)+\epsilon  \zeta  (\zeta -\theta -1)+\zeta \right] \nonumber \\
&+& \tpsi ^5 \left(\epsilon ^2-\zeta \right)^2(\zeta -\theta )^2 
\end{eqnarray*}
and
\begin{eqnarray*} 
D_1&=&\tilh (\epsilon -1) \tpsi ^5 \left(\epsilon ^2-\zeta \right) (\zeta -\theta ) \left[\epsilon ^2+\zeta  (\zeta -\theta -1)\right] \nonumber \\
&+& \tilh (\epsilon -1)^3 \epsilon ^2 \tpsi  (2 \epsilon -\theta -1) \nonumber \\
&-&\tilh (\epsilon -1)\tpsi ^3 \nonumber \\
&\times & \left[ \epsilon ^4 (3 \zeta -2 \theta )+2 \epsilon ^3 \left(\zeta ^2-2 \zeta  (\theta +1)+\theta ^2+\theta \right)\right] \nonumber \\
&+&\tilh (\epsilon -1)\tpsi ^3 \nonumber \\
&\times &\epsilon ^2 \left[ (\zeta +1) \theta ^2-2\zeta (\zeta +2)\theta +\zeta  (\zeta  (\zeta +4)-2)+\theta \right] \nonumber \\
&-&\tilh (\epsilon -1)\tpsi ^3 \nonumber \\
&\times & 2\epsilon \zeta^2+\zeta ^2 (\zeta -\theta -1)\nonumber \\
&-&\tpsi ^6 \left(\epsilon ^2-\zeta \right)^2 (\zeta -\theta )^2-(\epsilon -1)^4 \epsilon ^2 \nonumber \\
&+&\epsilon \tpsi ^4 \left(3 \epsilon ^3-\epsilon ^2 (2 \zeta +3)+\epsilon +\zeta ^2\right) (\zeta -\theta )^2 \nonumber \\
&-&(\epsilon -1)^3 \tpsi ^2 \left(\epsilon ^3-2 \epsilon ^2 \zeta +\zeta^2\right). 
\end{eqnarray*}
Notably, RHS of Eq. \refe{A:Q} depends on $s$ only through $\tpsi=\psis$ and $\tpsi_1 = \psisf$, while depends on $k$ only through $\tilh = \hk$. 
We obtained explicit form of the sharp propagator of the process containing two-step memory. Soft propagator can be simply obrained using (\ref{rown:Pks}).

\section{Variance and velocity autocorrelation function}

The results obtained in this appendix are based on explicit form of the process propagator derived in Appendix \ref{section:AppendA} for the two-step memory.
This propagator was derived in Appendix \ref{section:AppendA} from the recursive equation \refe{A:rekur} according to a conditional dependence of three consecutive changes in the price given by equation \refe{two:h}. An exact closed form of the propagator $\tQ {}{k,s}$ was given there by Eq. \refe{A:Q}.
Using Eq. \refe{rown:Pks}, the soft propagator $\tilde{P}(k,s)$ was derived. Next, the Laplace transform of the time-dependent process variance was derived 
\bee
\tilde{m}_2 (s)=-\left.\frac{\partial^2 \tilde{P}(k,s)}{\partial k^2}\right|_{k=0}=-\frac{\mu_2}{s^2 \tsr} \frac{N_2}{D_2}
\eee
where, obviously, $\tpsi=\psis$, while remaining quantities are parameters independent on $s$ and 
\bee
N_2&=&\tpsi ^3 \left(\zeta - \epsilon ^2 \right) (\zeta -\theta )+(\epsilon -1)^2 \tpsi  \left(2 \epsilon ^2-\zeta \right)+\nonumber \\
&&+\epsilon  \tpsi ^2 (2 (\epsilon-1) \epsilon -\zeta +1) (\zeta -\theta )-(\epsilon -1)^2 \epsilon \nonumber\\
D_2 &=&\tpsi ^3 \left(\zeta - \epsilon ^2 \right) (\zeta -\theta )+\epsilon  \tpsi ^2 (-2 \epsilon +\zeta +1) (\zeta -\theta)+\nonumber \\
&&+(\epsilon -1)^2 \zeta  \tpsi +(\epsilon -1)^2 \epsilon 
\eee
Finally, we obtain the Laplace transform of the velocity autocorrelation function of the process where explicit dependence on parameters $\zeta$ and $\theta$ is well seen.
\bee
C(s,\zeta,\theta)=-\frac{\mu_2}{2 \tsr} \frac{N_2}{D_2}
\eee
We remind that for $ \zeta = \theta = \epsilon ^2$ model with two-step memory well reproduces the results of that with one-step memory. Then,
\bee
C(s,\zeta=\epsilon^2,\theta=\epsilon^2)&=&\frac{\mu_2}{2 \tsr}\frac{1-\epsilon \psis}{1+\epsilon \psis},
\eee
which is the result derived already in \cite{gubiec2}.

For the autocorrelation function of the process including two-step memory, one just accept the value of the parameter $\theta = 0$, what leads to (\ref{two:tC})

\bibliography{my}
\bibliographystyle{unsrt}

\end{document}